\documentstyle[editedvolume,epsfig]{crckapb}

\begin{opening}
\title{THE CHANDRA X-RAY OBSERVATORY (CXO)}
\subtitle{An Overview}

\author{MARTIN C. WEISSKOPF}
\institute{NASA/MSFC\\
           Marshall Space Flight Center, AL 35801, USA}

\end{opening}

\runningtitle{THE CHANDRA X-RAY OBSERVATORY (CXO) }

\begin{document}

\section{Introduction}

Significant advances in science inevitably occur when the state of the art in instrumentation improves. NASA's newest Great Observatory, the Chandra X-Ray Observatory (CXO) -- formally known as the Advanced X-Ray Astrophysics Facility (AXAF) -- launched on July 23, 1999 and represents such an advance. The CXO is designed to study the x-ray emission from all categories of astronomical objects from normal stars to quasars. Observations with CXO will therefore obviously enhance our understanding of neutron stars and black holes. 

CXO has broad scientific objectives and an outstanding capability to provide high-resolution ($\leq$ 0.5-arcsec) imaging, spectrometric imaging and high resolution dispersive spectroscopy over the energy band from 0.1 to 10 keV. CXO, together with ESA's XMM, the Japanese-American Astro-E and ultimately the international Spectrum-X mission lead by Russia, will usher in a new age in x-ray astronomy and high-energy astrophysics.

NASA's Marshall Space Flight Center (MSFC) manages the Chandra Project, with scientific and technical support from the Smithsonian Astrophysical Observatory (SAO). TRW's Space and Electronics Group was the prime contractor and provided overall systems engineering and integration. Hughes Danbury Optical Systems (HDOS), now Raytheon Optical Systems Incorporated, figured and polished the x-ray optics; Optical Coating Laboratory Incorporated (OCLI) coated the polished optics with iridium; and Eastman Kodak Company (EKC) mounted and aligned the optics and provided the optical bench. Ball Aerospace \& Technologies was responsible for the Science Instrument Module (SIM) and the CCD-based aspect camera for target acquisition and aspect determination. The scientific instruments, discussed in some detail below, comprise two sets of objective transmission gratings that can be inserted just behind the 10-m-focal-length x-ray optics, and two sets of focal-plane imaging detectors that can be positioned by the SIM's translation table. 

The fully deployed CXO, shown schematically in Figure \ref{fig:cxo}, is 13.8-m  long, with a 19.5-m-long solar-array wingspan. The on-orbit mass is about 4500-kg. CXO was placed in a highly elliptical orbit with a 140,000-km apogee and 10,000-km perigee by the Space Shuttle Columbia, Boeing's Inertial Upper Stage, and Chandra's own integral propulsion system. Figure \ref{fig:cxo_launch} shows photos of the payload and the sequence of events through the deployment from the Shuttle.  This particular launch gained some additional notoriety due to the Commander's (Colonel Eileen Collins) gender.

\begin{figure}
\caption{Schematic of the CXO fully deployed.}
\label{fig:cxo}
\end{figure}

\begin{figure}
\caption{Chandra launch sequence.The IUS is still attached to the CXO in these photos.}
\label{fig:cxo_launch}
\end{figure}

\section{The X-ray Optics}
The heart of the observatory is, of course, the x-ray telescope. Grazing-incidence optics function because x rays reflect efficiently if the angle between the incident ray and the reflecting surface is less than the critical angle. This critical grazing angle is approximately $10^{-2}(2\rho)^{1/2}/E$, where $\rho$ is the density in g-cm$^{-3}$ and E is the photon energy in keV. Thus, higher energy telescopes must have dense optical coatings (iridium, platinum, gold, etc.) and smaller grazing angles. The x-ray optical elements for Chandra and similar telescopes resemble shallow angle cones, and two reflections are required to provide good imaging over a useful field of view; the first CXO surface is a paraboloid and the second a hyperboloid -- the classic Wolter-1 design. The collecting area is increased by nesting concentric mirror pairs, all having the same focal length. The wall thickness of the inner elements limit the number of pairs, and designs have tended to fall into two classes: Those with relatively thick walls achieve stability, hence angular resolution, at the expense of collecting area; those with very thin walls maximize collecting area but sacrifice angular resolution. NASA's Einstein Observatory (1978), the German ROSAT (1990), and the CXO optics are examples of the high-resolution designs, while the Japanese-American ASCA (1993) and European XMM mirrors are examples of emphasis upon large collecting area.

The mirror design for CXO includes eight optical elements comprising four paraboloid/hyperboloid pairs which have a common ten meter focal length, element lengths of 0.83-m, diameters of 0.63, 0.85, 0.97, and 1.2-m, and wall thickness between 16-mm and 24-mm. Zerodur, a glassy ceramic, from Schott was selected for the optical element material because of its low coefficient of thermal expansion and previously demonstrated capability (ROSAT) of permitting very smooth polished surfaces.

Figure \ref{fig:polish_h1} shows the largest optical element being ground at HDOS. Final polishing was performed with a large lap designed to reduce surface roughness without introducing unacceptable lower frequency figure errors. The resulting rms surface roughness over the central 90\% of the elements varied between 1.85 and 3.44 $\AA$ in the 1 to 1000-mm$^{-1}$ band; this excellent surface smoothness enhances the encircled energy performance at higher energies by minimizing scattering. 

\begin{figure}
\caption{The largest optical element being ground.}
\label{fig:polish_h1}
\end{figure}

The mirror elements were coated at OCLI by sputtering with iridium over a binding layer of chromium. OCLI performed verification runs with surrogates before each coating of flight glass; these surrogates included witness samples. The x-ray reflectivities of the witness flats were measured at SAO to confirm that the expected densities were achieved. The last cleaning of the mirrors occurred at OCLI prior to coating, and stringent contamination controls were begun at that time because both molecular and particulate contamination have adverse impacts on the calibration and the x-ray performance. Figure \ref{fig:coated_optic} shows the smallest paraboloid in the OCLI handling fixture after being coated.

\begin{figure}
\caption{The smallest parabaloid after coating.}
\label{fig:coated_optic}
\end{figure}

The final alignment and assembly of the mirror elements into the High Resolution Mirror Assembly (HRMA) was done at, and by, EKC. The completed mirror element support structure is shown in Figure \ref{fig:hrma_fixture}. Each mirror element was bonded near its mid-station to flexures previously attached to the carbon fiber composite mirror support sleeves. The four support sleeves and associated flexures for the paraboloids can be seen near the top of the figure, and those for the outer hyperboloid appear at the bottom. The mount holds more than 1000 kg of optics to sub-arcsecond precision.

\begin{figure}
\caption{The fixture to which the eight optical elements were mounted.}
\label{fig:hrma_fixture}
\end{figure}

The mirror alignment was performed with the optical axis vertical in a clean and environmentally controlled tower. The mirror elements were supported to approximate a gravity-free and strain-free state, positioned, and then bonded to the flexures. A photograph taking during the assembly and alignment process is shown in Figure \ref{fig:hrma_tower_ekc}. Despite the huge mass of the system and the stringent environmental controls, the heat produced by a 50 watt light bulb at the top of the facility caused some alignment anomalies until detected and resolved.

\begin{figure}
\caption{A photograph of the HRMA during assembly and alignment at EKC.}
\label{fig:hrma_tower_ekc}
\end{figure}

The HRMA was taken to MSFC for pre-launch x-ray calibration (see O'Dell and Weisskopf (1998) and references therein) in the fall of 1996, and then to TRW for integration into the spacecraft. Testing at MSFC took place in the X-Ray Calibration Facility (XRCF),  shown in Figure \ref{fig:xrcf_air}. The calibration facility has a number of x-ray source and detector systems and continues to be used for x-ray tests of developmental optics for such programs as Constellation-X. Details concerning the XRCF may be found in Weisskopf and O'Dell (1997) and references therein.  

X-ray testing demonstrated that the CXO mirrors are indeed the largest high-resolution X-ray optics ever made; the nominal effective area (based on the ground calibrations) is shown as a function of energy in the left panel of Figure \ref{fig:ea_ee}, along with those of their Einstein and ROSAT predecessors. The CXO areas are about a factor of four greater than the Einstein mirrors. The effective areas of CXO and ROSAT are comparable at low energies because the somewhat smaller ROSAT mirrors have larger grazing angles; the smaller grazing angles of CXO yield more throughput at higher energies. The fraction of the incident energy included in the core of the expected CXO response to 1.49-keV x rays is shown as a function of image radius in the right panel of Figure \ref{fig:ea_ee} including early in-flight data. The responses of the Einstein and ROSAT mirrors also are shown. The improvement within 0.5-arcsec is dramatic, although it is important to note that the ROSAT mirrors bettered their specification and were well matched to the principal detector for that mission. The excellent surface smoothness achieved for the CXO (and ROSAT) mirrors result in a very modest variation of the  performance as a function of energy; this reduces the uncertainties which accrue from using calibration data to infer properties of sources with different spectra, and improves the precision of the many experiments to be performed. 

\begin{figure}
\caption{An aerial view of the X-ray Calibration Facility at MSFC.}
\label{fig:xrcf_air}
\end{figure}

\begin{figure}
\caption{Effective area and encircled energy comparisons.}
\label{fig:ea_ee}
\end{figure}

\section{The Instruments}
CXO has two focal plane instruments -- the High-Resolution Camera (HRC) and the Advanced CCD Imaging Spectrometer (ACIS). Each of these instruments, in turn, has two detectors, one optimized for direct imaging of x rays that pass through the optics and the other optimized for imaging x rays that are dispersed by the objective transmission gratings when the latter are commanded into place directly behind the HRMA. Each focal-plane detector operates in essentially photon counting mode and has low internal background. A slide mechanism is utilized to place the appropriate instrument at the focus of the telescope. Provision for focus adjustment is also present. 

\subsection{The Focal Plane Instruments}
The HRC was produced at SAO; Dr. S. Murray is the Principal Investigator. The HRC-I is a large-format, 100-mm-square microchannel plate, coated with a cesium iodide photocathode to improve x-ray response. A conventional cross-grid charge detector reads out the photo-induced charge cloud and the electronics determine an arrival time to 16$\mu$s, and the position with a resolution of about 18 $\mu$m or 0.37 arcsec. The spectroscopy readout detector (HRC-S) is a 300-mm x 30-mm, 3-section microchannel plate. Sectioning allowed the 2 outside sections to be tilted in order to conform more closely to the Rowland circle that includes the low-energy gratings.

The ACIS has 2 charge coupled-device (CCD) detector arrays: ACIS-I is optimized for high-resolution spectrometric imaging; ACIS-S is optimized for readout of the high-energy transmission gratings, although these functions are not mutually exclusive. Prof. G. Garmire of the Pennsylvania State University is the Principal Investigator. The Massachusetts Institute of Technology's Center for Space Research, in collaboration with Lincoln Laboratories, developed the detector system and manufactured the CCDs; Lockheed-Martin integrated the instrument. Stray visible light is shielded by means of baffles and an optical blocking filter (about 1500-$\AA$ aluminum on 2000-$\AA$ polyimide). The ACIS-I is a 2x2 array of CCDs. The 4 CCDs tilt slightly toward the optics to conform more closely to the focal surface. Each CCD has 1024 x 1024 pixels of 24-$\mu$m (0.5-arcsec) size. The ACIS-S is a 1x6 array with each chip tilted slightly to conform to the Rowland circle and includes two back-illuminated CCDs, one of which is at the best focus position. The back-illuminated devices cover a broader bandwidth than the front-illuminated chips and, under certain circumstances, may be the best choice for high-resolution, spectrometric imaging.

\subsection{The Transmission Gratings}

Both sets of objective transmission gratings consist of hundreds of co-aligned facets mounted to supporting structures on 4 annuli (one for each of the four co-aligned mirror pairs) to intercept the x rays exiting the HRMA. In order to optimize the energy resolution, the grating support structure holds the facets close to the Rowland toroid that intercepts the focal plane. The two sets of transmission gratings, attached to the mounting structure are shown in Figure \ref{fig:gratings_mounted}. 

\begin{figure}
\caption{The LETG and HETG being attached to the spacecraft mounting structure.}
\label{fig:gratings_mounted}
\end{figure}

The Low-Energy Transmission Grating (LETG) provides high-resolution spectroscopy at the lower end of the CXO energy range. Dr. A Brinkman, of the Space Research Organization of the Netherlands, is the Principal Investigator. The LETG was developed in collaboration with the Max Planck Institut f\"ur Extraterrestische Physik, Garching. The LETG has 540 1.6-cm diameter grating facets, 3 per grating module. Ultraviolet contact lithography was used to produce an integrated all-gold facet bonded to a stainless-steel facet ring. An individual facet has 0.43-$\mu$m-thick gold grating bars with 50\% filling factor and 9920-$\AA$ period, resulting in 1.15-$\AA$/mm dispersion. The HRC-S is the primary LETG readout.

The High-Energy Transmission Grating (HETG) provides high-resolution spectroscopy at the higher end of the CXO energy range. Prof. C. Canizares of the Massachusetts Institute of Technology Center for Space Research is the Principal Investigator. This group developed the instrument in collaboration with MIT's Nanostructures Laboratory. The HETG has 336 2.5-cm square grating facets. Microlithographic fabrication using laser interference patterns was used to produce the facets, which consist of gold grating bars with 50\% filling factor on a polyimide substrate. The HETG uses gratings with 2 different periods which are oriented to slightly different dispersion directions, forming a shallow "X" image on the readout detector as shown in Figure \ref{fig:hetg_x}.  

\begin{figure}
\caption{The first observation of Capella with the HETG/ACIS-S combination.}
\label{fig:hetg_x}
\end{figure}

The Medium-Energy Gratings (MEG) have 0.40-$\mu$m-thick gold bars on 0.50-$\mu$m-thick polyimide with 4000-$\AA$ period, producing 2.85-$\AA$/mm dispersion, and are placed behind the outer two CXO mirrors. The High-Energy gratings (HEG), placed behind the inner two CXO mirror pairs are 0.70-$\mu$m -thick-gold bars on 1.0-$\mu$m-thick polyimide with 2000-$\AA$ period, resulting in 5.7-$\AA$/mm dispersion. The ACIS-S is the primary readout for the HETG.

\section{Science with CXO }
Given the superb capabilities of the optics and associated instrumentation, the scientific possibilities are almost incredible. The most exciting investigations will no doubt result from the unexpected discoveries that the improved sensitivity, angular resolution, and energy resolution produces. The potential of Chandra is illustrated in Figure \ref{fig:casa_firstlight} which shows the official "first-light" image. The target was the supernova remnant Cas A. This image, based on only a few thousand seconds observing time, was taken with the back-illuminated chip at the best focus position on the ACIS-S. We see, for the first time, that there is a compact, x-ray emitting object at the center of the 300-year-old remnant. Studies are underway to establish that the positional coincidence is no accident and to determine the nature of the compact object, possibly the long-sought after neutron star or black hole.

\begin{figure}
\caption{The official CXO first light image -- the supernova remnant CasA.}
\label{fig:casa_firstlight}
\end{figure}

Perhaps the neutron star - black hole connection and the utility of CXO are best illustrated by an early observation of the Crab Nebula and pulsar taken as part of the HETG calibration. During grating observations, one also obtains an undispersed (zero order) image. The image quality is essentially that of the HRMA/detector combination and not broadened by the insertion of the grating. Figure \ref{fig:crab_chandra_b} shows the zeroth-order image of the Crab Nebula. There are numerous new features, especially the inner ellipse with its bright knots. The pulsar itself is so bright that the central region is "piled up" to the point that there are no data -- hence the "hole" in the image. Pile-up also is present in the data from the nebula  and the study of the spectral dependence of these features ought to be the subject of a future ASI. The ubiquity of the jet phenomenon clearly points to the importance of angular momentum as a physical key and critical parameter towards unlocking secrets to all or part of the emission mechanisms. 

\begin{figure}
\caption{The Chandra zeroth order image of the Crab Nebula.}
\label{fig:crab_chandra_b}
\end{figure}

The capability to perform meaningful, high-spectral-resolution observations with the gratings is illustrated in Figure \ref{fig:hetg_Fe_XVII}, which shows a portion of the x-ray spectrum from Capella around the Fe-L complex. The red is a HEG spectrum and the green spectrum was produced by the MEG. Observations such as these -- with CXO, XMM, and Astro-E -- will be at the center of new developments in astrophysics in the next century.  

\begin{figure}
\caption{HEG and MEG spectra of Capella.}
\label{fig:hetg_Fe_XVII}
\end{figure}

\section{Conclusion}
The Chandra X-Ray Observatory will have a profound influence on astronomy and astrophysics. The Observatory is open to use by scientists throughout the world who successfully propose specific investigations. Data will be available through the Chandra X-ray Center (CXC), directed by Dr. H. Tananbaum, and located at the SAO.

\section*{Acknowledgements}
I would like to thank the many members of the Chandra team, especially Steve O'Dell, Leon Van Speybroeck, Harvey Tananbaum, Steve Murray, Gordon Garmire, Claude Canizares, and Bert Brinkman.

\appendix
\section{AXAF web sites}
The following lists several Chandra-related sites on the World-Wide Web (WWW). Most sites are cross-linked to one another. Often you will find that these contain the best and most recent sources of detailed information; hence, the minimal number of entries in the bibliography.

\begin{description}

\item[\verb|http://asc.harvard.edu/|] Chandra X-Ray Center (CXC), operated for NASA by the Smithsonian Astrophysical Observatory (SAO).

\item[\verb|http://wwwastro.msfc.nasa.gov/|] Chandra Project Science, at the NASA Marshall Space Flight Center (MSFC).

\item[\verb|http://hea-www.harvard.edu/MST/|] AXAF Mission Support Team (MST), at the Smithsonian Astrophysical Observatory (SAO).

\item[\verb|http://hea-www.harvard.edu/HRC/|] AXAF High-Resolution Camera (HRC) team, at the Smithsonian Astrophysical Observatory (SAO).

\item[\verb|http://www.astro.psu.edu/xray/axaf/axaf.html|] Advanced CCD Imaging Spectrometer (ACIS) team at the Pennsylvania State University (PSU).

\item[\verb|http://acis.mit.edu/|] Advanced CCD Imaging Spectrometer (ACIS) team at the Massachusetts Institute of Technology (MIT).

\item[\verb|http://www.sron.nl/|] Chandra Low-Energy Transmission Grating (LETG) team at Space Research Organisation Netherlands (SRON).

\item[\verb|http://www.rosat.mpe-garching.mpg.de/axaf/|] Chandra Low-Energy Transmission Grating (LETG) team at the Max-Planck Institut f\"ur extraterrestrische Physik (MPE).

\item[\verb|http://space.mit.edu/HETG/|] Chandra High-Energy Transmission Grating (HETG) team, at the Massachusetts Institute of Technology (MIT).

\end{description}


\begin{thebibliography}{}
\bibitem[]{} O'Dell, S.L. and Weisskopf, M.C. (1998) {\it Proc SPIE}, {\bf 3444}, 2
\bibitem[]{} Weisskopf, M.C. and O'Dell, S.L. (1997) {\it Proc SPIE}, {\bf 3113}, 2

\end{thebibliography}
\end{document}